# Magnetism in (Ga,Mn)As Thin Films With $T_C$ Up To 173K

K.Y. Wang, R.P. Campion, K.W. Edmonds, M. Sawicki[1], T. Dietl[1], C.T. Foxon, and B.L. Gallagher

*School of Physics and Astronomy, University of Nottingham, Nottingham NG7 2RD, United Kingdom*
[1]*Institute of Physics, Polish Academy of Sciences, al. Lotników 32/46, PL-02668 Warszawa, Poland*

**Abstract.** We have investigated the magnetic properties of (Ga,Mn)As thin films with Mn concentration between 1 and 9%. Ferromagnetic transition temperatures $T_C$ of up to 173K are observed. The results are compared to the predictions of the Zener mean-field theory. We find no evidence of a fundamental limit to $T_C$.

## INTRODUCTION

The III-V dilute magnetic semiconductor (Ga,Mn)As offers new possibilities for the effective integration of ferromagnetic properties into semiconductor heterostructures [1]. The substitutionally incorporated Mn is an acceptor, providing valence band holes that mediate a ferromagnetic ordering of the local moments. Most of the key magnetic properties of this material are remarkably well described by a Zener-type mean-field model [2], which, in its simplest form, predicts that the ferromagnetic transition temperature $T_C$ varies as $x \cdot p^{1/3}$, where $x$ is the concentration of Mn and $p$ is the concentration of holes. Until recently, the highest reported $T_C$ remained steadfastly at 110 K, prompting some speculation that this could be a fundamental limit [3]. However, this 'limit' has since proved to be due to a purely technological matter of incorporating concentrations of Mn well above the equilibrium solubility, which necessitates low-temperature growth. This can lead to a high density of defects, the most important of which are are As antisites, $As_{Ga}$, and interstitial Mn, $Mn_I$, both compensating double donors in (Ga,Mn)As. $Mn_I$ may also couple antiferromagnetically to the substitutional $Mn_{Ga}$, further reducing $T_C$. Steps to minimize defect densities by careful control of growth and post-growth annealing conditions have recently resulted in substantial improvements in $T_C$ [4-6].

Here, we describe the preparation and characterisation of a series of high-$T_C$ (Ga,Mn)As films, with Mn concentrations between 1 and 9%.

## EXPERIMENTAL DETAILS

The (Ga,Mn)As layers are grown on low-temperature GaAs buffer layers on semi-insulating GaAs(001) by molecular beam epitaxy [7]. The Mn concentration is determined from the Mn/Ga flux ratio, which was calibrated by SIMS measurements on 1μm thick samples grown under the same conditions. The material quality is found to be strongly dependent on the growth temperature and As/Ga flux ratios. The former is chosen to be the highest possible while maintaining 2D growth (≈190-300°C, depending on Mn flux), as monitored by RHEED. The films were grown under relatively low As flux, in order to minimize the concentration of compensating $As_{Ga}$. The use of $As_2$ as a source rather than $As_4$ also appears to favour lower $As_{Ga}$ concentrations [7].

The films are annealed in air at 190°C for several hours. This is a well-established procedure for removing compensating $Mn_I$ from the layers [3-6], by diffusion to the free surface followed by passivation by e.g. oxidation. For this reason, the highest $T_C$ values are obtained for layers of thickness 50 nm or lower, without protective capping layers. The optimum anneal time for a given thickness is obtained by monitoring the electrical resistance during annealing, which effectively gives an *in-situ* measurement of the hole density [4].

$T_C$ is obtained both from anomalous Hall measurements using Arrott plots, and from the temperature-dependence of the remnant magnetization measured by SQUID. The hole density is estimated from Hall and magnetoresistance measurements at 0.3 K and 16 T, using a fitting procedure to separate normal and anomalous terms.

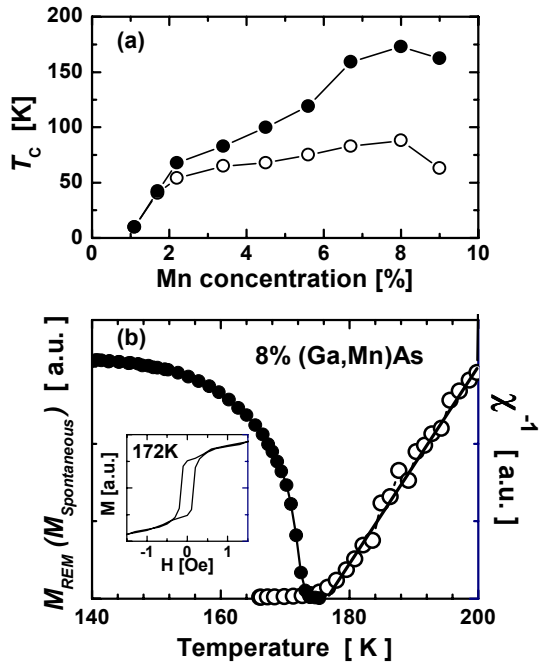

**FIGURE 1.** (a) Ferromagnetic transition temperatures versus total Mn concentration for as-grown (open) and annealed (filled) (Ga,Mn)As thin films; (b) temperature-dependence of remnant magnetization and inverse paramagnetic susceptibility for the *x*=0.08 sample; inset: hysteresis loop for the same sample at 172 K.

## RESULTS AND DISCUSSION

Figure 1a shows the ferromagnetic transition temperatures as a function of the total Mn concentration in as-grown and annealed (Ga,Mn)As single layers of thickness 25-50 nm. Annealing clearly has a pronounced effect on $T_C$, especially at high Mn concentrations. The measured resistivity and hole density show similar trends. This effect has been shown to be due to the removal of compensating interstitial Mn from the layers [3-6].

The highest value obtained so far is 173 K, for a 25 nm thick sample with nominal 8% Mn. To our knowledge this is the highest value reported in (Ga,Mn)As single layers. The remnant magnetization and inverse paramagnetic susceptibility versus temperature for the 8% sample is shown in figure 1b. The inset shows clear ferromagnetic hysteresis at 172 K for this sample.

Even in the annealed samples, $T_C$ appears to rise significantly more slowly than the $x^{4/3}$ predicted by the two-band mean field model, in the absence of compensation. However, it is important to remember that the quoted Mn concentration includes compensating interstitial Mn as well as the magnetically active $Mn_{Ga}$. On annealing, the $Mn_I$ are fully removed from the lattice rather than moving onto substitutional sites [6]. The concentration of magnetically active $Mn_{Ga}$ will therefore be smaller than the total Mn concentration measured by SIMS on unannealed samples. From hole density measurements before and after annealing, we can estimate the concentrations of $Mn_I$ and $Mn_{Ga}$. We find that at high concentrations, 20-30 % of the Mn occupies interstitial sites in the as-grown films, and also that the hole density is comparable to the $Mn_{Ga}$ density in most of the annealed films [8]. The mean-field-predicted values of $T_C$ for the hole and substitutional Mn concentrations we obtain are in good quantitative agreement with the experimental results [8]. This indicates that incorporating higher concentrations of substitutional Mn will produce further increases of $T_C$.

## ACKNOWLEDGMENT

This work has been supported by EC FENIKS and CELDIS projects, UK EPSRC, and Polish KBN.